\documentclass[final, review, 10pt]{article}

\usepackage{longtable}
\usepackage{amssymb}
\usepackage{amsmath}
\usepackage{amsfonts}
\usepackage{graphicx}
\usepackage{color}
\usepackage{epstopdf}
\usepackage{url}
\usepackage{hyperref}
\usepackage{float}
\usepackage{titlesec}
\usepackage[top=1in, bottom=1.25in, right=1.025in, left=1.25in]{geometry}
\usepackage{framed}
\usepackage{multicol}
\usepackage{footnote}
\usepackage{threeparttable}
\usepackage{longtable}
\usepackage{array}
\usepackage{ragged2e}
\usepackage{filecontents}
\usepackage{setspace}
\usepackage{hyperref}
\usepackage{caption}
\usepackage{subcaption}
\makesavenoteenv{tabular}
\makesavenoteenv{table}
\usepackage{threeparttable}
\usepackage[toc]{appendix}
\usepackage{breqn}
\usepackage{ragged2e}
\usepackage{epstopdf}
\usepackage{subcaption}
\usepackage{authblk}

\usepackage{soul}
\usepackage{lipsum}                     
\usepackage{xargs}                      
\usepackage[pdftex,dvipsnames]{xcolor}
\usepackage[colorinlistoftodos,prependcaption,textsize=small]{todonotes}
\newcommandx{\Rnotes}[2][1=]{\todo[linecolor=red,backgroundcolor=red!24,bordercolor=red,#1]{#2}}

\definecolor{mycol2}{rgb}{0.75, 0.98, 0.75}

\justifying
\begin{document}

\doublespacing
\begin{titlepage}
\title{Enhanced Equivalent Circuit Model for High Current Discharge of Lithium-Ion Batteries with Application to Electric Vertical Takeoff and Landing Aircraft}
\author[1]{Alireza Goshtasbi \thanks{Corresponding author: alireza.goshtasbi@jobyaviation.com}}
\author[1]{Ruxiu Zhao}
\author[1]{Ruiting Wang}
\author[1]{Sangwoo Han}
\author[1]{Wenting Ma}
\author[1]{Jeremy Neubauer}
\affil[1]{Powertrain and Electronics Department, Joby Aviation, San Carlos, CA}

\date{}
\maketitle
\vspace{-10mm}

\begin{abstract}

	Conventional battery equivalent circuit models (ECMs) have limited capability to predict performance at high discharge rates, where lithium depleted regions may develop and cause a sudden exponential drop in the cell’s terminal voltage. Having accurate predictions of performance under such conditions is necessary for electric vertical takeoff and landing (eVTOL) aircraft applications, where high discharge currents can be required during fault scenarios and the inability to provide these currents can be safety-critical. To address this challenge, we utilize data-driven modeling methods to derive a parsimonious addition to a conventional ECM that can capture the observed rapid voltage drop with only one additional state. We also provide a detailed method for identifying the resulting model parameters, including an extensive characterization data set along with a well-regularized objective function formulation. The model is validated against a novel data set of over 150 flights encompassing a wide array of conditions for an eVTOL aircraft using an application-specific and safety-relevant reserve duration metric for quantifying accuracy. The model is shown to predict the landing hover capability with an error mean and standard deviation of 2.9 and 6.2 seconds, respectively, defining the model’s ability to capture the cell voltage behavior under high discharge currents.

\end{abstract}


\end{titlepage}
\newpage

\section{Introduction} \label{Sec:Intro}

Lithium-ion batteries used in electric vertical takeoff and landing (eVTOL) applications must provide both high power and energy density, while ensuring fault tolerance \cite{sripad2021promise, fredericks2018performance, yang2021challenges}. In a hover where one of multiple battery packs are offline due to a fault, discharge currents up to and exceeding 8C may be required of the battery cells. Inability to deliver this current in its entirety may result in the rapid loss of altitude. Preventing this requires high-rate battery hardware; however, as the available energy of every battery is finite, a hardware solution alone is not sufficient (i.e., a pilot can always attempt to fly beyond the capability of the battery). Thus, accurate high-rate battery state estimation and prediction tools are critical to planning flights within hardware capabilities and delivering a safe eVTOL product.

Battery management systems (BMS) commonly rely on models to provide state estimation and performance prediction functions \cite{lin2019modeling}. Models used for such purposes can vary in fidelity depending on the available computational resources and the accuracy requirements of the application at hand; ranging from high fidelity physics-based models to empirical equivalent circuit representations. While physics-based models offer better insights into the true response of internal states, parameter identifiability and state observability often pose a challenge and mandate the use of model reduction methods \cite{berliner2021nonlinear, park2018optimal, moura2015estimation}. Moreover, real-time implementation of such models may not be possible in resource constrained applications \cite{sturm2019suitability}. Such limitations have promoted the adoption of empirical models, such as equivalent circuit models (ECM), for online performance prediction and state estimation.

In its simplest form, an ECM consists of several parallel resistor-capacitor (RC) pairs to capture dynamic voltage response in series with a resistor to capture the ohmic response \cite{nejad2016systematic}. Properly parameterized, this type of ECM is effective at low C-rates, but prediction accuracy can degrade significantly at higher currents. This has prompted the development of ECMs that attempt to account for solid phase diffusion effects using fractional order dynamics to represent the constant phase diffusion elements in the time domain \cite{alavi2015time}, or by distinguishing between surface and bulk concentrations and introducing an empirical relation between the two \cite{biju2023battx, zheng2019accurate, braun2022state, li2022novel, ouyang2014enhancing, zhuo2023diffusion}. More recently, a receding horizon approach has been proposed to capture the diffusion effects using a discrete convolution operator \cite{li2021discrete, li2023interpretable}. This approach appears promising, but obtaining good approximations requires long horizons that in turn increase the dimension of the model. Nonlinear system identification methods have also been used, wherein the linear ECM dynamics are combined with static nonlinearities to improve model prediction accuracy \cite{tian2020nonlinear, widanage2016design}.

These methods appear to perform well in automotive and other applications. However, lithium-ion cells employed for the eVTOL application can be subject to extreme discharge rates. The combined effects of solid phase diffusivity and lithium depletion (LD) in the electrolyte under such high discharge rates can lead to rapid cell voltage decay \cite{lain2021understanding, fuller1994simulation}. This behavior is different from the open-circuit-voltage (OCV) knee at low states of charge (SOC), and may happen at moderate to high SOCs where the diffusion limited C-rate (DLC) is exceeded \cite{heubner2020DLC, zuo2022DLC}. A low-dimensional ECM that captures both electrolyte and solid phase diffusion limitations and replicates this behavior has remained elusive. Note that most state estimation methods such as Kalman filters scale as $\mathcal{O}(n^3)$, where $n$ is the dimension of the state-space. Therefore, keeping the model dimension low is critical to enabling pack-level state estimation.

Another challenge in using ECMs is model parameterization. While ECMs are often identifiable (under mild assumptions) thanks to their simple structure \cite{alavi2016identifiability}, their parameters typically must vary with SOC, temperature, and often current to provide satisfactory accuracy under diverse operating conditions. Some studies have related ECM parameters to physical quantities to assist identification and provide qualitative insights \cite{hariharan2013nonlinear, zhang2017novel, geng2021bridging, pang2021comprehensive}, but this may sacrifice accuracy. Others have fit constant parameters to short duration charge/discharge pulse data at different conditions, then created parameter lookup tables from or fit functions to this data \cite{lin2014lumped, bruch2021novel}. This approach is not suitable for modeling the combined solid phase diffusion and LD response, because the sustained high C-rate conditions that lead to the accelerated voltage drop are often accompanied by considerable changes in the SOC and temperature that invalidate the constant parameter assumption.

Finally, eVTOL applications require accurate energy indications to aid pilots in their landing decisions and satisfy regulatory requirements. Predictions of the available reserve time at the landing location is crucial in this regard. The commonly used model validation metrics that are based on voltage prediction errors are not good proxies for such energy indications; under LD conditions at high discharge rates small errors in estimating the DLC can lead to significantly large voltage errors. Thus, available reserve time prediction error is more appropriate for model evaluations in eVTOL applications.

In this work, we use data-driven modeling techniques to develop an enhanced ECM formulation (hereafter referred to as LD-ECM) that accurately captures the LD behavior (where we use the LD denomination loosely to refer to all diffusion limited processes in a cell) and is suitable to use for eVTOL battery state estimation and performance prediction. We further develop and demonstrate improved model parameterization methods suitable to this model using extensive test data from large-format commercial pouch cells, and validate the resultant model to a large set of eVTOL flight profiles on the metric of reserve duration accuracy. The rest of the paper is organized as follows; Section \ref{Sec:Models} describes the LD-ECM derivation and Section \ref{Sec:ParameterID} details the parameter identification approach and the test data used for building and refining the model. Validation results on eVTOL flight profiles are given in Section \ref{Sec:Validation} before summary and conclusions in Section \ref{Sec:Conclusions}. A complete list of abbreviations is provided in \mbox{\ref{Sec:nomenclature}}.


\section{Enhanced ECM Derivation} \label{Sec:Models}

\subsection{Conventional ECM}
\begin{figure}[t!]
	\begin{center}
		\includegraphics[width=0.9\textwidth]{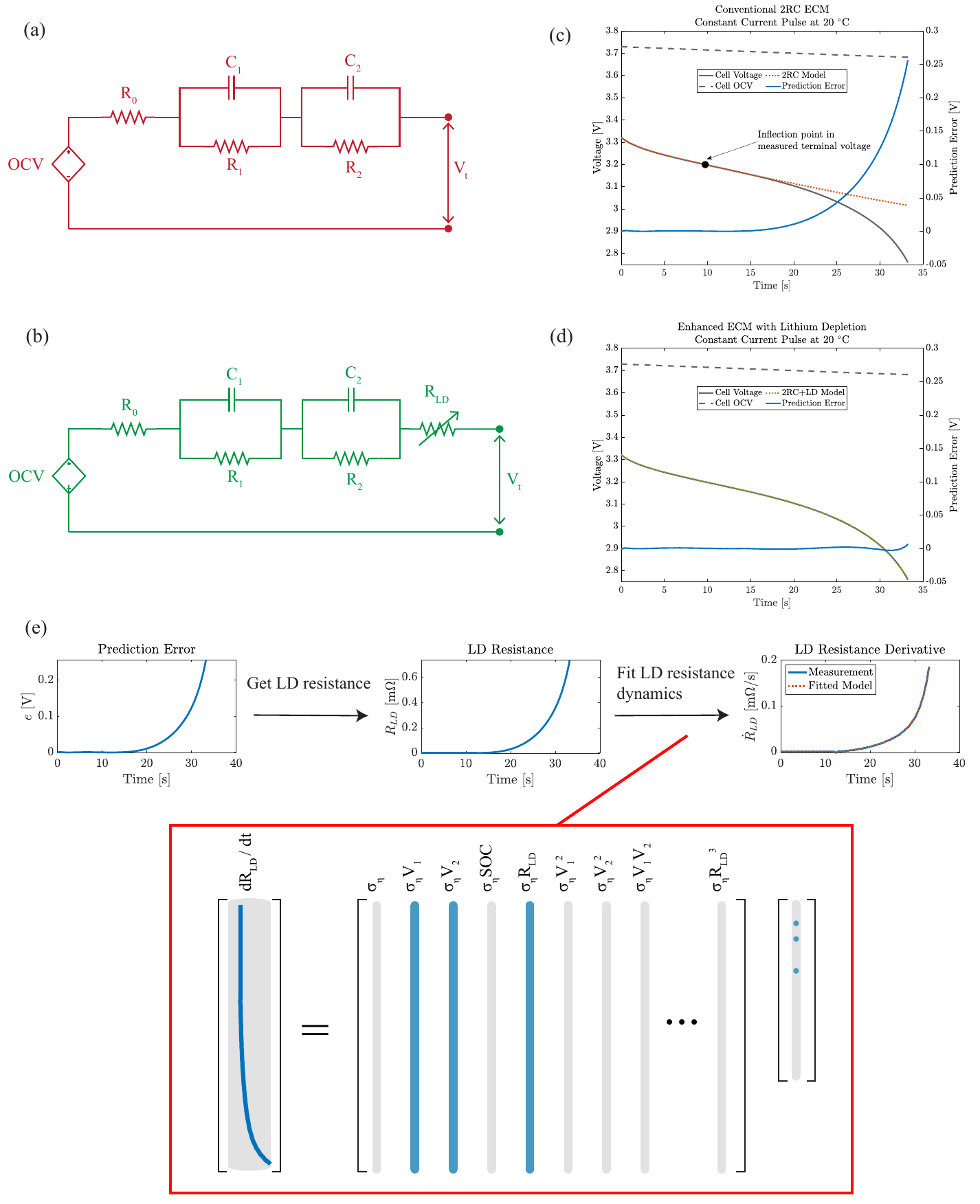}
		\vspace{-0.3cm}
		\caption{Schematic of (a) conventional 2RC ECM, (b) proposed LD enhanced ECM, and model prediction in presence of LD behavior for (c) conventional 2RC ECM, and (d) LD-ECM, and (e) discrepancy modeling framework from \cite{brunton2016SINDy} used to learn the LD dynamics.}
		\vspace{-0.7cm}
		\label{Fig:RcComparison}
	\end{center}
\end{figure}

A traditional 2RC ECM, as shown in Fig. \ref{Fig:RcComparison}(a), is typically sufficient to capture slower than 1 Hz dynamics inside a lithium-ion cell \cite{nejad2016systematic, hentunen2014time}, thus serving as the starting point for this work. The state and output equations for this 2RC ECM are given by:
\begin{align}
	\frac{dSOC}{dt} &= \frac{I}{36 Q} \label{Eq:SOC},\\
	\frac{d{V}_1}{dt} &= \frac{-V_1}{\tau_1}  + \frac{IR_1}{\tau_1},\\
	\frac{d{V}_2}{dt} &= \frac{-V_2}{\tau_2}  + \frac{IR_2}{\tau_2},\\
	V_{t, 2RC} &= OCV(SOC, T) + V_1 + V_2 + I R_0,
\end{align}
where $SOC$ is the cell state of charge in percent, $I$ is the current (negative for discharge), $Q$ is the cell capacity in Ampere hours, $V_i$ is the voltage across the $i$-th RC pair, $\tau_i=R_i C_i$ is the corresponding time constant, and $OCV$ is the SOC and temperature dependent open-circuit voltage. Observed OCV hysteresis is minor in our selected cell, and therefore neglected for simplicity. 

To highlight the shortcomings of this conventional 2RC ECM at capturing high-rate LD behavior, we present an example high C-rate pulse data in Fig. \ref{Fig:RcComparison}(c), in which the terminal voltage is concave as it approaches the lower limit of 2.75V. Assuming (for illustration purposes only) constant circuit parameters through the relatively short pulse, and recognizing the OCV in this region of SOC is nearly linear, it is apparent that this over-damped second order ECM will only predict convex voltage curves.  An alternative model architecture is required to recreate such behavior.

\subsection{Discrepancy Modeling for Enhanced ECM}
Discrepancy modeling seeks to identify a sparse dynamical system that can explain the observed prediction errors \cite{kaheman2019discrepancyModeling, ebers2022discrepancy}. This identified dynamical system is then added to the original model. Sparsity is a key feature to avoid overfitting and only finding the simplest model that can explain the observed behavior. 

The method fits the prediction error derivative using the fewest terms from a library of regressors predefined by the user \cite{brunton2016SINDy}. For the ECM in consideration, this means finding a sparse representation for the right hand side of the following equation:
\begin{align*}
	\frac{de}{dt} = f\left(e, V_1, V_2, SOC, I\right),  \label{Eq:edot}
\end{align*}
where $e$ denotes the prediction error. In our experience, having the right terms in the regressor library is critical to successful application of the sparse regression framework. Indeed, an efficient representation may not be viable in the wrong coordinates \cite{champion2019coordinates}. After exploring multiple avenues for finding the simplest model, we make three design choices prior to applying the discrepancy modeling framework.
\begin{enumerate}
	\item The observed error can be explained by a dynamic series resistance:
	\begin{align}
		e = I R_{LD}. \nonumber
	\end{align} 
	This assumption is derived from the observation that $\rm Li^+$ depletion leads to a sudden increase in ohmic resistance in the electrolyte phase. The electrolyte ionic conductivity is a function of $\rm Li^+$ concentration \cite{landesfeind2019temperature}. As $\rm Li^+$ concentration goes to zero in the positive electrode domain under high C-rates, the electrolyte ohmic resistance increases accordingly. As the depletion region grows, so does the cell's ohmic resistance. This assumption is validated later when we measure the cell ohmic resistance for high current discharge pulses.
	
	\item LD resistance growth has a triggering function, $\sigma_{\eta}$, driven by the total cell overpotential ($V_1 + V_2$):
	\begin{align*}
		\sigma_{\eta} = \frac{1}{2}\left[1 + \tanh\left(\frac{-(V_1 + V_2) - \eta_{th}}{\delta_{\eta}}\right)\right],
	\end{align*}
	where $\eta_{th}$ denotes the threshold overpotential above which LD growth is triggered and $\delta_{\eta}$ is a smoothing parameter determining the transition width for the sigmoid function ($\delta_{\eta}=4$mV  throughout this work \footnote{This value was chosen empirically based on initial numerical experimentation. A smaller transition width would make the sigmoid function change more abruptly. This in turn can make parameter identification harder, since the activation function would have a non-zero gradient in a smaller range of overpotentials. On the other hand, larger values may lead to earlier and unintended activation of the LD resistance growth.}). The threshold overpotential can be approximated based on the total overpotential of the cell at the inflection point that is identified in Fig. \ref{Fig:RcComparison}(c). Finding the inflection point requires calculating the second derivative of the measured voltage, which makes it sensitive to measurement noise. We use a total variation denoising algorithm \cite{selesnick2012tvd} to attenuate the noise before differentiation.
	
	This function is designed to mimic the electrochemical behavior of the cell when approaching and exceeding the DLC. The DLC depends on the initial $\rm Li^+$ concentration in the electrolyte \cite{heubner2020DLC}, which in turn depends on the load history the cell has been subjected to. Therefore, the DLC at a given temperature and SOC could be lower in presence of concentration gradients compared to an equilibrium state. In the ECM, deviation from the equilibrium state is captured by the RC voltages (i.e., $V_1$ and $V_2$). Hence, the total predicted overpotential in the cell ($\eta = -V_1 - V_2$) is used as the activation signal. 
	
	\item We separate self-growth and forced growth terms, i.e., we do not include any cross terms between $R_{LD}$ and the other independent variables ($SOC$, $V_1$, and $V_2$). This is solely to allow for more efficient model representation, as including the cross terms was found to lead to more complicated models without improved accuracy.
	
\end{enumerate}

The LD resistance growth (error) dynamics can thus be written as:
\begin{align}
	\frac{d{R}_{LD}}{dt} = \sigma_{\eta} \times \left[f_1\left(V_1, V_2, SOC\right) + f_2\left(R_{LD}\right)\right],
\end{align}
where the first term on the right hand side ($f_1$) determines the forced growth rate of the LD resistance, while the second term ($f_2$) governs its self growth rate. Note that when starting from zero initial condition, LD resistance only starts growing when $\sigma_{\eta}$ is activated. We can now apply the sparse regression framework to identify the terms in $f_1$ and $f_2$. To that end, we consider monomials of up to third degree, including cross terms between the input arguments of $f_1$. Following steps similar to those described by Brunton et al. \cite{brunton2016SINDy}, we arrive at a simple linear formulation:
\begin{align}
	\frac{d{R}_{LD}}{dt} = \sigma_{\eta} \times \left[ -\theta_{\eta}\left(V_1 + V_2\right) + \theta_{R} R_{LD}\right], \label{Eq:Rdot}
\end{align}
where $\theta_{\eta}$ and $\theta_{R}$ are the forced and self growth rate scalar parameters identified experimentally. To ensure monotonic behavior and avoid over-fitting, we force both $\theta_{R}$ and $\theta_{\eta}$ to be non-negative. This means that the above equation leads to an unstable mode in the model to represent the observed voltage drop. However, the empirically observed phenomenon is reversible and the LD resistance relaxes upon a reduction in load. This is accommodated by further modifying the above equation as follows:
\begin{align}
	\frac{d{R}_{LD}}{dt} = \sigma_{\eta} \times \left[ -\theta_{\eta}\left(V_1 + V_2\right) + \theta_{R} R_{LD}\right] - (1 - \sigma_{\eta})\times \frac{R_{LD}}{\tau_{LD}}, \label{Eq:Rdot_w_relaxation}
\end{align}
where $\tau_{LD}$ is the relaxation time constant for the LD resistance. The LD-ECM architecture is illustrated in Fig. \ref{Fig:RcComparison}(b) and the resulting fit to the same measurements is given in Fig. \ref{Fig:RcComparison}(d) showing significant improvement over the conventional 2RC ECM. The discrepancy modeling framework used to arrive at the above formulation is illustrated in Fig. \ref{Fig:RcComparison}(e).

Lastly, the addition of the LD resistance modifies the terminal voltage equation as follows:
\begin{align}
	V_{t, LD-ECM} &= OCV(SOC, T) + V_1 + V_2 + I \times \left(R_0 + R_{LD}\right).
\end{align}
Compared to the conventional 2RC ECM, the LD-ECM has one additional dynamics state, i.e., $R_{LD}$, and three additional parameters, namely, $\eta_{th}$, $\theta_{\eta}$, and $\theta_{R}$, thereby allowing for a low dimensional representation of the observed LD behavior at high C-rates. Overall, the LD-ECM model can be represented in the simple state-space format as follows:
\begin{align}
	\frac{d}{dt} \begin{bmatrix}
		SOC\\ 
		V_1\\ 
		V_2\\
		R_{LD} 
	\end{bmatrix} = 
	\begin{bmatrix}
		0&  0&  0& 0\\ 
		0&  \frac{-1}{\tau_1}&  0& 0 \\ 
		0& 0 & \frac{-1}{\tau_2} & 0\\ 
		0& -\sigma_{\eta}\theta_{\eta} & -\sigma_{\eta}\theta_{\eta} & \sigma_{\eta}\theta_{R}  - \frac{1-\sigma_{\eta}}{\tau_{LD}}
	\end{bmatrix}
	 \begin{bmatrix}
		SOC\\ 
		V_1\\ 
		V_2\\
		R_{LD} 
	\end{bmatrix} + 
	 \begin{bmatrix}
		\frac{1}{36Q}\\ 
		\frac{R_1}{\tau_1}\\ 
		\frac{R_2}{\tau_2}\\
		0 
	\end{bmatrix} I.
\end{align}
The compact state-space formulation simplifies use of the model for state estimation tasks. Nonetheless, the unstable eigenvalue of the system (recall that $\theta_R\ge0$) amplifies voltage sensitivity to parameter perturbations and makes parameter identification challenging.

\subsection{Model Parameterization}

We adopt the typical linear parameter varying (LPV) approach to ECMs where each parameter is a function of SOC, temperature \cite{lin2014lumped, bruch2021novel}, and current \cite{hentunen2014time, karimi2023equivalent}. The current dependency is not desirable due to the additional dimensionality and the algebraic loop created by a load-dependent $R_0$ when power is the model input, but our experimentation with removing the current dependency from all or some of the parameters resulted in significantly degraded model quality. The model parameters are thus given by:
\begin{align}
	\theta_{j, k} = h_j(SOC_k, T_k, I_k, \boldsymbol{\alpha}_j), \label{Eq:h_j}
\end{align}
where $\theta_j$ is the LD-ECM parameter of interest (either one of $R_0, R_1, R_2, \tau_1, \tau_2, \theta_{\eta}, \theta_{R}, \eta_{th}$), $k$ is the time index, and $h_j$ is the function representing the variations in parameter $\theta_j$, which is itself parameterized by $\boldsymbol{\alpha}_j$. All $\theta_{j}$'s must be non-negative in our formulation.

We evaluate using both lookup tables (LUT) with linear interpolation and functional representations for $h_j$'s. For the LUT formulation, we use a uniform grid in all three dimensions per Table \ref{Table:LutGrid}. For the functional representation, we explore multivariate polynomials of varying degrees as well as rational functions. In the rest of the paper we focus on the LUT representation, as it was found to result in better model performance. Further details about the functional representation are provided in \ref{Sec:Functional_Forms}.

\begin{table}[t!]
	\caption{Grid points for the discharge LUTs}
	\vspace{-4mm}
	\label{Table:LutGrid}
	\begin{center}
		\begin{tabular}{l  c }
			\hline
			Variable [Units] & Grid Points \\
			\hline \vspace{2mm}
			$SOC$ $\rm [\%]$ & 0-100\% in 5\% increments  \\\vspace{1mm}
			$T$ $\rm [^\circ C]$  & 10-65 $\rm ^\circ C$ in 5$\rm ^\circ C$  increments \\\vspace{1mm}
			$I$ $\rm [A]$  & $-\left[0.1, 0.5, 1.0, 2.0, 3.0, 4.0, 5.0, 6.0, 7.0, 8.0\right]\times Q$\\  
			\hline		
		\end{tabular}
	\end{center}
\end{table}

\section{Methods: Parameter Identification} \label{Sec:ParameterID}

\subsection{Characterization Experimental Design}

\begin{figure}[t!]
	\begin{center}
		\includegraphics[width=0.80\textwidth]{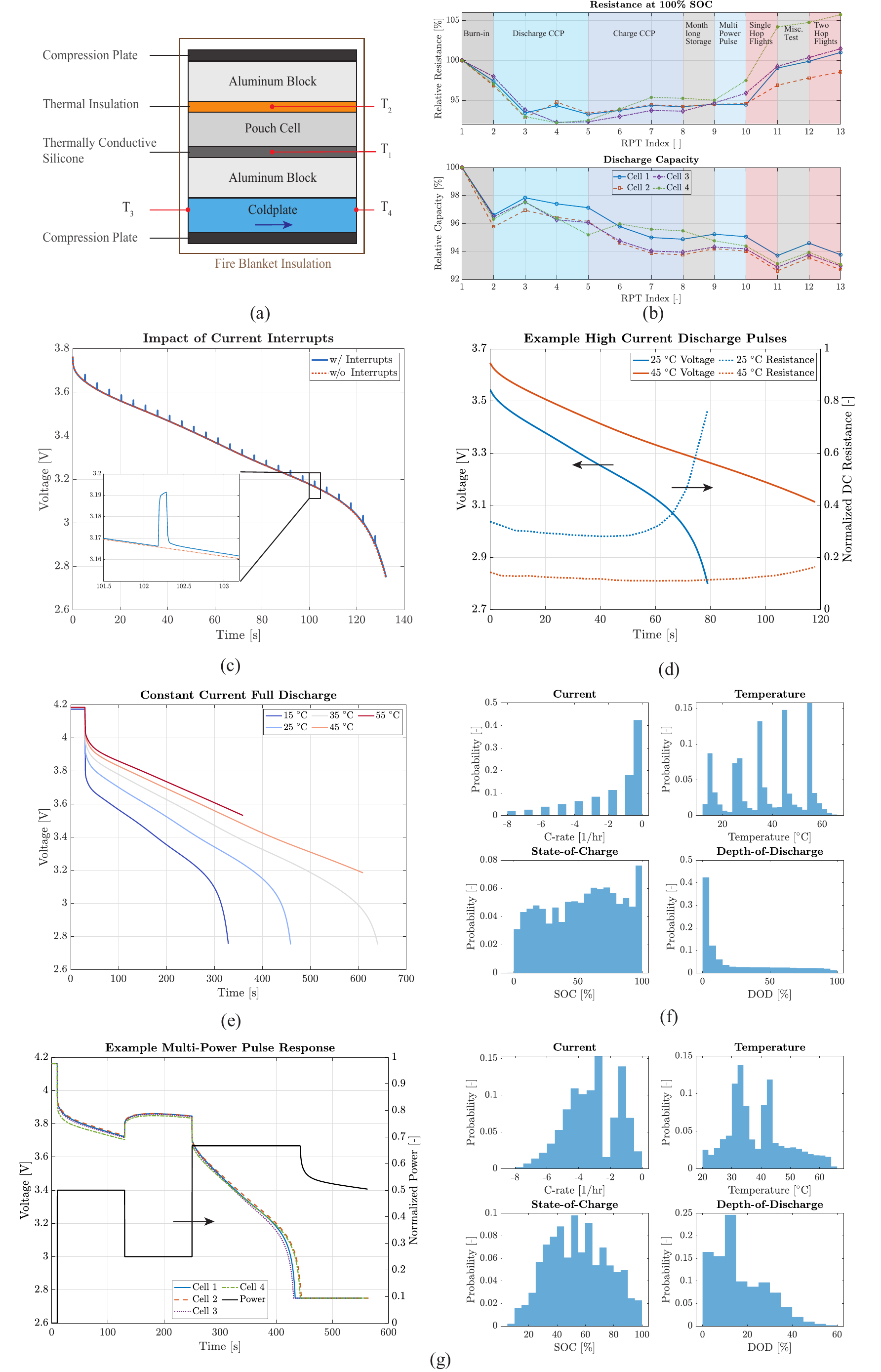}
		\vspace{-0.3cm}
		\caption{Characterization test data: (a) test setup, (b) RPT data for all cells used in test campaign, (c) high C-rate CCP pulse showing minimal impact of current interrupts, (d) example high C-rate discharge pulses and their corresponding DC resistance estimated with the interrupt method, (e) example full discharge data at different temperatures, (f) histograms of data coverage for all constant current (CCP and FD) tests, and (g) MPP example and data coverage histograms.}
		\vspace{-0.7cm}
		\label{Fig:CharacterizationTests}
	\end{center}
\end{figure}

All test data are collected on large format commercial stacked pouch cells with graphite negative and NMC positive electrode. The pouches have an area to thickness ratio of $5.5 \times 10^{-4}$ and are tested under an applied pressure of $\sim$70 kPa. Four cells from the same batch are used throughout the testing campaign. All tests are conducted in a blast chamber and temperature control is achieved by actively running temperature-controlled coolant (de-ionized water) through a coldplate that mates one surface of the cell, as shown in Fig. \ref{Fig:CharacterizationTests}(a). The temperature reading $T_2$ is used as the cell temperature in all model evaluations.

All cells are initially cycled 110 times using a fast charging protocol and a simplified eVTOL flight discharge profile that achieves a deep depth of discharge such that the cells are past their initial rapid degradation phase prior to characterization testing. The cells’ state-of-health (SOH) is periodically checked using a standard reference performance test protocol (RPT) that includes both capacity and resistance checks. The annotated RPT results for the entire test campaign are shown in Fig. \ref{Fig:CharacterizationTests}(b). The apparent difference between the degradation of the cells is due to several factors, including inherent manufacturing variation in large format pouch cells, potential differences, albeit small, between thermal boundary conditions experienced by the cells, as well as the cells being subject to different loads for certain parts of the testing campaign (between RPTs 2 and 8).

\begin{table}[t!]
	\caption{DOE for constant current pulse characterization (CCP)}
	\vspace{-4mm}
	\label{Table:DOE}
	\begin{center}
		\begin{tabular}{l  c }
			\hline
			Variable [Units] & Grid Points \\
			\hline \vspace{2mm}
			$OCV$ $\rm [V]$ & 3.28-4.18 V in 0.06 V increments  \\\vspace{1mm}
			$T$ $\rm [^\circ C]$  & 15-55 $\rm ^\circ C$ in 10$\rm ^\circ C$  increments \\\vspace{1mm}
			$I$ $\rm [A]$  & $-\left[0.1, 0.5, 1.0, 2.0, 3.0, 4.0, 5.0, 6.0, 7.0, 8.0\right]\times Q$\\  
			\hline		
		\end{tabular}
	\end{center}
\end{table}

The first characterization test type is a set of constant current pulses (CCP) based on a full factorial design of experiment (DOE) with OCV (our proxy for SOC), temperature, and current as the experimental variables, as described in Table \ref{Table:DOE}. This DOE yields a total of 800 CCP discharges. Each discharge pulse is 2 minutes long unless the cell hits the lower voltage limit of 2.75 V or an upper temperature limit of 65 $\rm ^\circ C$. Due to the large size of the DOE, each cell in the batch was used to cover a quarter of the design.

Discharge pulses at and above 5C include small current interrupts of -0.2C magnitude for a duration of 0.2 seconds every 5 seconds to directly measure the change in ohmic resistance of the cell throughout the pulse as shown in Fig. \ref{Fig:CharacterizationTests}(c). This same figure also illustrates the near identical performance of the same cell tested to this high current pulse without the interrupts. Thus, once the ohmic resistance is estimated for each pulse, the impact of interrupts is removed from the voltage data to ensure data integrity when down-sampled to 1 second intervals for fitting.

Fig. \ref{Fig:CharacterizationTests}(d) shows two high C-rate CCP pulses at 25 and 45 $\rm ^\circ C$ and the corresponding resistances measured using the current interrupts. The voltage curves shown have been processed as noted above to remove the effect of the interrupts. While both pulses start at $\sim$67\% SOC, only the 25 $\rm ^\circ C$ exhibits the accelerated voltage decay that is the sign of LD and a corresponding exponential increase in ohmic resistance. The measured resistance of 45 $\rm ^\circ C$ pulse remains nearly constant. This serves as an empirical confirmation of the assumption that LD exhibits an increase in a cell’s ohmic resistance.

The second characterization test type is a set of full discharge (FD) constant current pulses from top of charge to the minimum voltage. These tests are conducted at the same temperatures as in the CCP set, but only for C-rates up to 6C. An example of this data set is shown in Fig. \ref{Fig:CharacterizationTests}(e) for a single C-rate.

While the CCP and FD data sets provide good coverage of the SOC, temperature, and current conditions (as shown in Fig. \ref{Fig:CharacterizationTests}(f)), they are all constant current and always start from rest. Therefore, a third characterization test is introduced. This test type is a multi-power pulse (MPP) consisting of three sequential constant power discharges. The first two steps last for 2 min each. The third step continues until a cell limit is reached. A test is terminated immediately if a temperature limit is encountered; if a voltage limit is encountered, discharge continues in a constant voltage operating mode. Power magnitudes are ordered medium, low, high and a total of 72 tests are conducted at temperatures of 20, 30, and 40 $\rm ^\circ C$. A sample profile and the resulting voltage response is shown in Fig. \ref{Fig:CharacterizationTests}(g). Data coverage histograms for the MPP tests are also provided in Fig. \ref{Fig:CharacterizationTests}(g).

\subsection{Identification Problem Formulation}

\subsubsection{Solution Strategy}
To map out parameter dependencies on operating conditions, often short pulses are used to identify ECM parameters at each SOC and temperature condition. The results are then utilized to fit constant model parameter at the specified conditions \cite{lin2014lumped}. This approach is appealing for its computational efficiency in the identification step, since each pulse can be fitted individually by solving a small optimization problem. Unfortunately, the constant parameter assumption breaks down for longer pulses or those at higher C-rates, since the SOC and temperature can vary meaningfully even within a 60 second time window. Hu et al. circumvented this issue by using subspace methods \cite{hu2011linear}. Here, we focus on identifying the full parameter function in one optimization using the entire characterization data set.

The large size of the data set and the high dimensional parameter space results in a large-scale optimization problem. To simplify the process, we first fit one separate parameter set to the CCP and FD data at each individual current level, reducing the number of decision variables in a single optimization problem by a factor of ten.  Subsequently, the complete large scale optimization  over all CCP, FD, and MPP data is conducted using the previous result as an initial guess.

\subsubsection{Optimization Formulation}

The parameter identification step seeks to find the $\boldsymbol{\alpha}$ vector for ECM parameters that best fits the characterization data. This is formulated as:
\begin{align}
	\min_{\boldsymbol{\alpha}} \, \quad & J_{\rm err}(\boldsymbol{\alpha}) + J_{\rm reg}(\boldsymbol{\alpha}) \label{Eq:Opt}\\
	{\rm s.t.} \quad & \underline{\boldsymbol{\alpha}} \le \boldsymbol{\alpha} \le \overline{\boldsymbol{\alpha}} \nonumber
\end{align}
where $\boldsymbol{\alpha}= \left[\boldsymbol{\alpha}_{R_0},\boldsymbol{\alpha}_{R_1}, \boldsymbol{\alpha}_{R_2}, \boldsymbol{\alpha}_{\tau_1}, \boldsymbol{\alpha}_{\tau_2}, \boldsymbol{\alpha}_{\theta_{\eta}},  \boldsymbol{\alpha}_{\theta_{R}}, \boldsymbol{\alpha}_{\eta_{th}}\right]^T$ is the vector of decision variables, $J_{\rm err}(\boldsymbol{\alpha})$ and $J_{\rm reg}(\boldsymbol{\alpha})$ are the prediction error and the regularization costs, respectively, while $\underline{\boldsymbol{\alpha}}$ and $\overline{\boldsymbol{\alpha}}$ denote the parameter lower and upper bounds that are selected based on cell testing data. The bounds are also used to enforce a sufficient time scale separation between the two RC pairs ($1.5 {\rm s} \le \tau_1 \le 10 {\rm s}$, $30 {\rm s} \le \tau_2 \le 150 {\rm s}$) to improve identifiability \cite{alavi2016identifiability, hentunen2014time}.

The prediction error cost in Eq. \ref{Eq:Opt} is given by:
\begin{align}
	J_{\rm err}(\boldsymbol{\alpha}) = \sum_{i=1} ^{n_{\rm pulse}}\Bigg[ &\frac{w_1}{\eta_{\rm C}} RMSE_{{\rm LD+RC}, \, i}(\boldsymbol{\alpha}) + \nonumber\\ 
	& \frac{w_2}{\eta_{\rm C}} RMSE_{{\rm diff}, \, i}(\boldsymbol{\alpha}) +  \nonumber \\
	&\frac{w_3}{\eta_{\rm C}} E_{{{\rm LD+RC}, \, n_t}, \, i}(\boldsymbol{\alpha}) + \nonumber \\ 
	& w_4 RMSE_{R_{\rm ohm}, \, i}(\boldsymbol{\alpha}) +  \nonumber \\
	&\frac{w_5}{\eta_{\rm C}} RMSE_{{\rm RC}, \, i}(\boldsymbol{\alpha}) + \nonumber \\
	& w_6 R_{{\rm LD}, \, {n_t}, \, i}(\boldsymbol{\alpha}) \Bigg]. \label{Eq:J_err}
\end{align}
All terms are evaluated individually for each pulse at a given current and the summation over all pulses is used in the objective function. The first term is the root mean squared error (RMSE) of the model. The second term is the RMSE of the voltage prediction error derivative and is used to force the predicted voltage curvature to more closely match the measurements. The third term additionally penalizes deviations in the last data point in each pulse, where the subscript $n_t$ is used to denote the last index in the pulse. This term is added to better match the time to min voltage, since that is an important criterion for available energy and power predictions. The fourth term is the RMSE of the ohmic resistance prediction error. Note that the predicted ohmic resistance in the model is $R_{\rm ohm, model} = R_0 + R_{LD}$ and ohmic resistance measurements are only available in the high C-rate pulse of the CCP data set. Hence, this term is set to zero when the measurements are not available. The fifth term is the RMSE of only the 2RC portion of the model (i.e., without LD resistance). This term is added to discourage the optimizer to use the LD resistance when the RC portion of the model is sufficient to fit the data. The last term penalizes the value of the LD resistance at the end of the pulse and serves as an additional guardrail against over-reliance on the LD term.

\begin{figure}[t!]
	\begin{center}
		\includegraphics[width=0.95\textwidth]{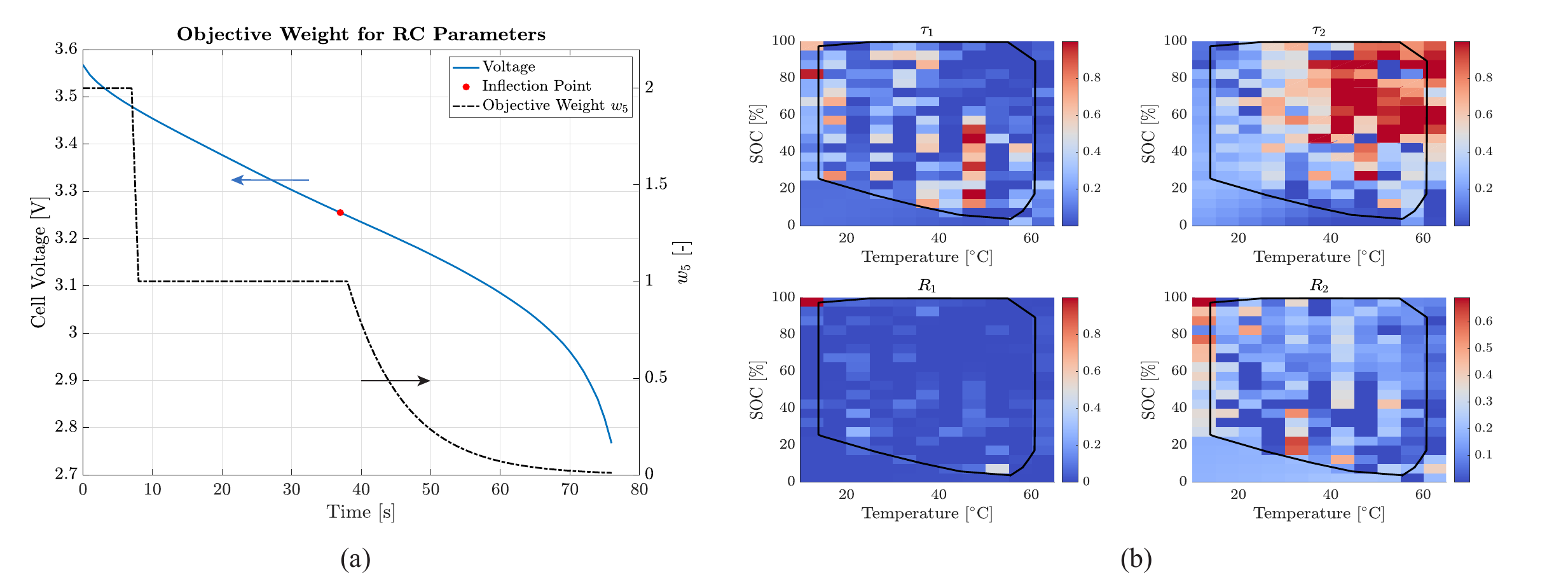}
		\vspace{-0.3cm}
		\caption{Creating the cost function for parameter identification: (a) the variable weight for RC-only error, and (b) example LUT results obtained with zero regularization, where the black boundary indicates the convex hull of the test data coverage in this example fitting case.}
		\vspace{-0.7cm}
		\label{Fig:CostFunction}
	\end{center}
\end{figure}

We note that some weights are divided by $\eta_{\rm C}$. This is a scaling factor that is equal to the average of measured overpotentials over all the pulses at a given C-rate. This is done to balance the error and regularization costs equally at different C-rates (since the overpotential will be smaller at lower currents and the error cost can be overwhelmed by the regularization cost without appropriate scaling). Lastly, all $w_i$'s except for $w_5$ are constants. $w_5$ is a time varying function defined separately for each pulse. An example time profile of $w_5$ is shown in Fig. \ref{Fig:CostFunction}(a), where it is seen to have three distinct regions; for the first 8 seconds, $w_5$ is given a higher value to deal with the data imbalance between the fast and slow dynamics \cite{hu2014two}. Subsequently, $w_5$ goes through a flat region, before starting an exponential decay. The final decaying behavior is added to allow the RC portion of the  model to deviate from the measurements past an inflection point, when one exists. This helps the optimizer to not bias the RC parameters to fit the LD behavior.

\subsubsection{Parameter Regularization: LUT}

With constant current characterization data, we can build the LUTs one current at a time. At a fixed current, each LUT has 252 grid points, leaving a total of 2016 decision variables for all 8 LD-ECM parameters (i.e., $\boldsymbol{\alpha}\in \mathbb{R}^{2016\times1}$). Because of this over-parameterized representation, care must be taken to ensure smoothness of the resulting LUTs and limit use of the LD resistance term where two RC pairs alone can sufficiently match the measurements. Indeed, our attempts at fitting the model without regularization guardrails resulted in accurate predictions on the training data. However, the resulting LUTs, shown in Fig. \ref{Fig:CostFunction}(b), lacked smoothness and led to large errors on the validation data sets. Thus, the following LUT regularization cost was developed:
\begin{align}
	J_{\rm reg}(\boldsymbol{\alpha}) = &\sum_{\substack{j \in \{R_0, R_1, R_2, \\ \tau_1, \tau_2, \theta_{\eta}, \theta_{R}, \eta_{th} \}}} \Bigg[ w_{{\rm reg, 1}, \, j} \left(\| \Delta_T \boldsymbol{\alpha}_j\|_2^2 + \| \Delta_{SOC}\boldsymbol{\alpha}_j\|_2^2\right) +\nonumber\\
	& w_{{\rm reg, 2}, \, j} \left(\| \Delta_T^2 \boldsymbol{\alpha}_j\|_2^2 + \| \Delta_{SOC}^2\boldsymbol{\alpha}_j\|_2^2\right) + w_{{\rm reg, I, 1}, \, j} \left(\| \Delta_I \boldsymbol{\alpha}_j\|_2^2 \right) \Bigg] +\nonumber\\
	&w_{\theta_{\eta}}\|\theta_{\eta}\|_1 + w_{\theta_{R}}\|\theta_{R}\|_1 - w_{\eta_{th}}\|\eta_{th}\|_1  - w_{{\tau_2}}\|\tau_{2}\|_1 , \label{Eq:J_reg_lut}
\end{align}
where the $\Delta_i$ operator is a difference operator applied in the $i$-th variable direction (to approximate the first derivative), $\Delta_i^2$ denotes two consecutive application of the operator in the respective direction (to approximate the second derivative), $\|.\|_k$ is the usual $k$-norm. The arguments inside the summation are penalizing the first and second derivatives of the parameters in both the SOC and temperature directions and the first derivative along the current direction. This ensures smoothness of the resulting parameter tables at a given current, and slowly varying parameters between different current levels. When fitting the LUTs, we start at the highest C-rate where the current regularization term is set to zero. For each subsequently lower C-rate fitting, the LUT values are regularized against their corresponding values identified at the previous current. Additional terms penalize the 1-norm of the LD parameters. This is done to minimize LD usage to only necessary conditions. Also, we note that all the weights are positive quantities. Therefore, the above formulation forces $\eta_{th}$ to be as high as possible, which has the effect of delaying the onset of LD, thereby minimizing its occurrence. A similar penalty is applied to $\tau_2$ (the bigger time constant in our model) to favor better matching the longer term behavior, when only shorter duration pulses are available for certain conditions (due to hitting a cell operational limit before the 2 minute pulse duration is over).

\subsubsection{Numerical Optimization}
For model training, the LD-ECM model is integrated using a forward Euler integration method with the time step of 1 second. Model integration is done using CasADi  \cite{Andersson2019} in Matlab to utilize its automatic differentiation for Hessian computations. For CCP fitting, we use a direct single shooting approach, where the entire trajectory is simulated in a recursive fashion and no additional decision variables are added to the original problem. However, for the FD data set  the length of the voltage sequence can exceed 3000 (e.g., for 1C discharge sampled at 1 second). For such long sequences, the Hessian expressions can become very cumbersome. We therefore use a partially constrained multiple shooting approach \cite{van2012tackling, retzler2022shooting}, where we break a long trajectory into 2 minute segments and introduce additional decision variables for states at the segmentation points. Constraints are also added to ensure state continuity at those points. This approach reduces the recursion length in each single shooting segment and simplifies the expressions in the Hessian. The resulting optimization problem in Eq. \ref{Eq:Opt} is then solved using IPOPT \cite{wachter2006implementation} with MA27 from the HSL libraries as the linear solver \cite{hsl2007collection}. In all cases, when fitting to constant current data at only one C-rate, we use the exact Hessian (as provided by CasADi). However, when fitting to the variable current MPP data set, the L-BFGS approximation \cite{nocedal1999numerical} is used for the Hessian. For LUT fitting, we start the highest C-rate pulse fitting with a uniform initial guess. But for subsequently lower C-rate pulses, we use the previous slice of the LUT to warm start the optimization. When fitting the functional forms described in \ref{Sec:Functional_Forms}, we warm start the optimization by initially fitting the function coefficients to the LUT data.

\subsection{Model Fitting Results}\label{Sec:FittingResults}

\subsubsection{Parameter LUTs and Performance on CCP, FD, and MPP}

\begin{figure}[t!]
	\begin{center}
		\includegraphics[width=0.9\textwidth]{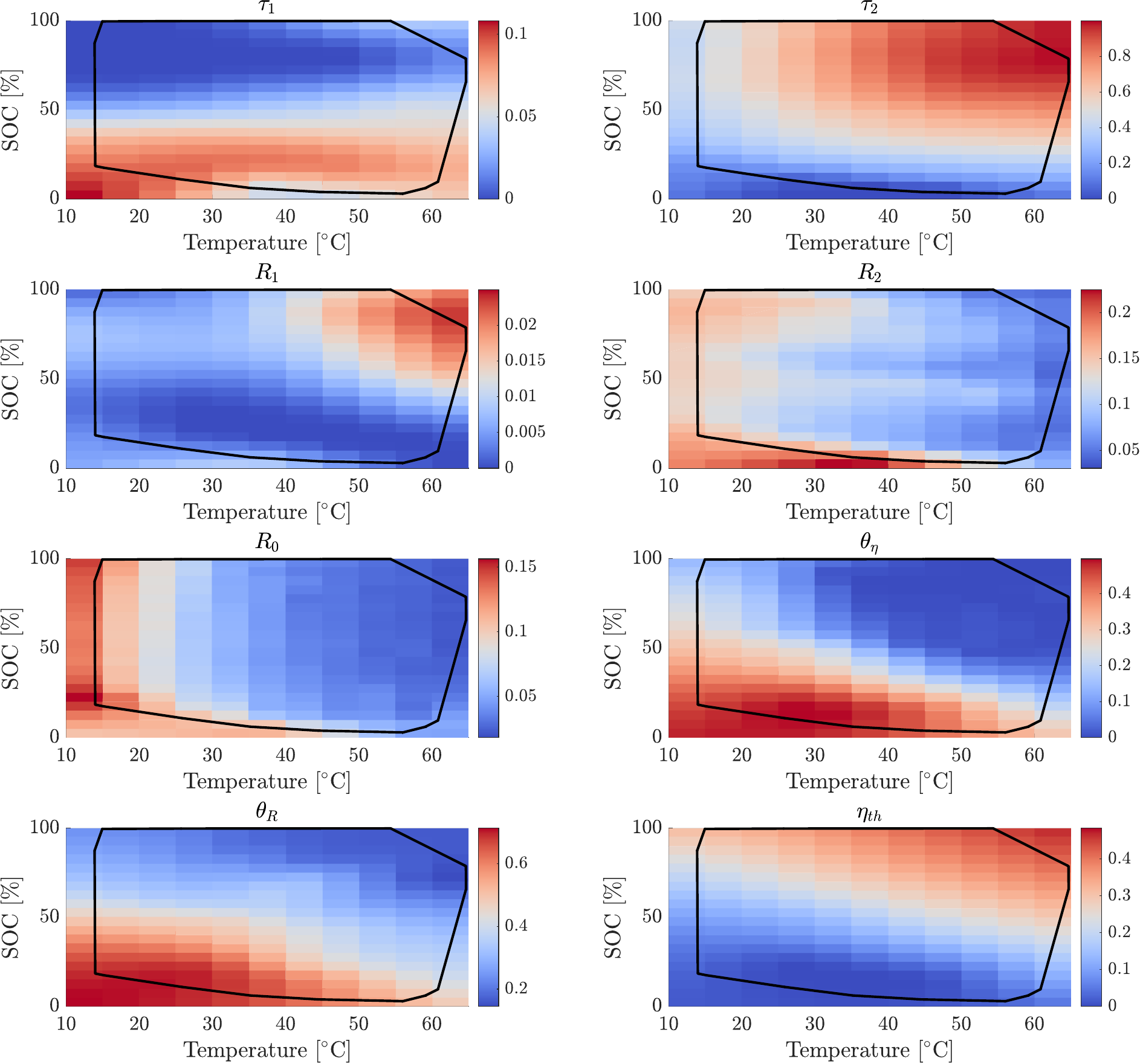}
		\vspace{-0.3cm}
		\caption{Fitted normalized lookup tables to 6C CCP data, where the black boundaries denote the convex hull of the CCP data coverage.}
		\vspace{-0.7cm}
		\label{Fig:LutAt6C}
	\end{center}
\end{figure}

\begin{figure}[t!]
	\begin{center}
		\includegraphics[width=0.9\textwidth]{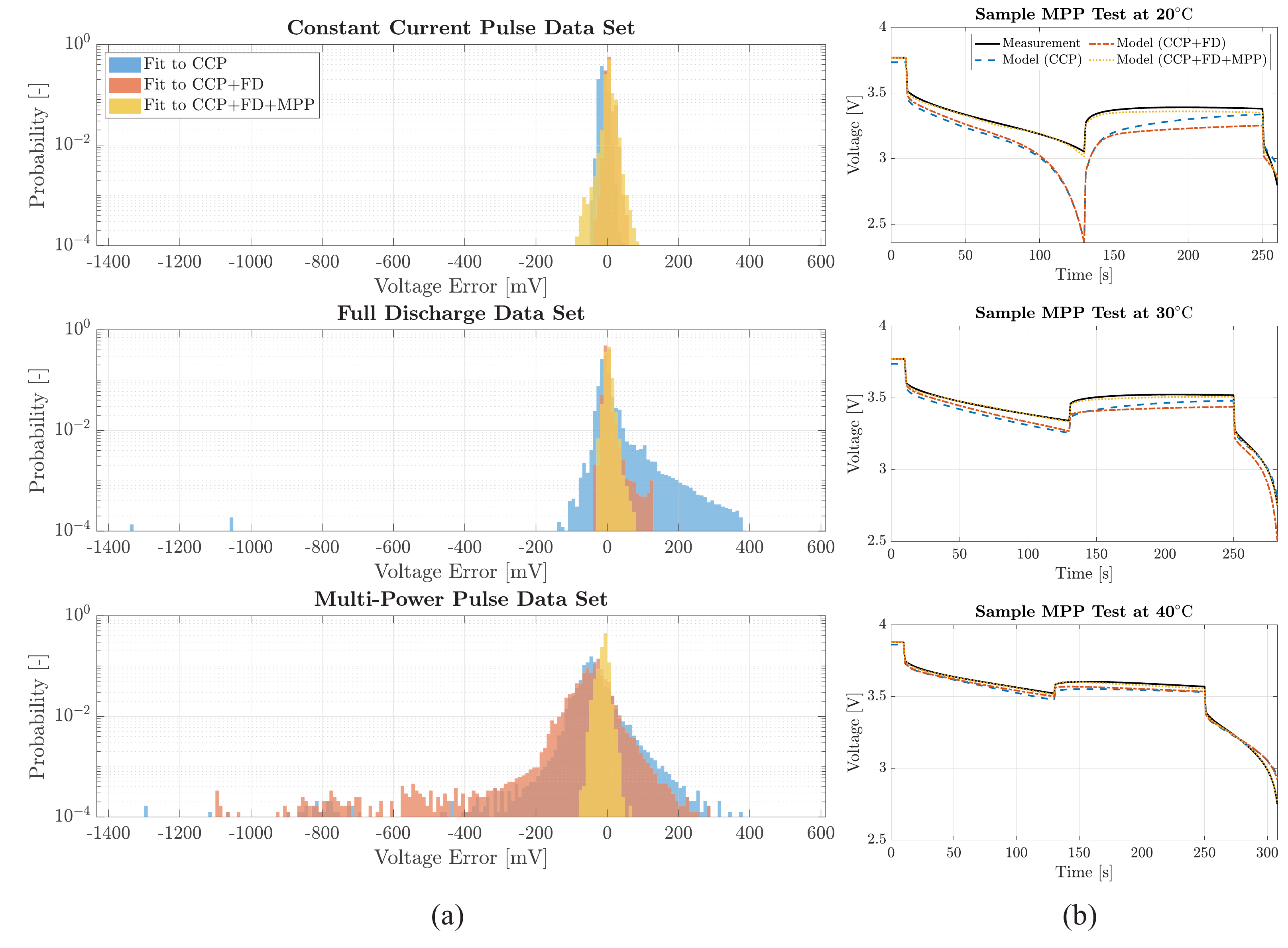}
		\vspace{-0.3cm}
		\caption{Impact of characterization data on model quality: (a) voltage prediction error histograms with different training data, where the $y$-axis is in logarithmic scale to better reveal the distribution tails, and (b) example MPP profiles and the corresponding model predictions when fitted to different characterization data.}
		\vspace{-0.7cm}
		\label{Fig:FitToMPP}
	\end{center}
\end{figure}

Fig. \ref{Fig:LutAt6C} shows exemplary parameter LUTs at 6C discharge current. Overall, the resulting parameter tables are sufficiently smooth thanks to the regularization terms in the objective function. The resistance tables $R_0$ and $R_2$ follow the expected trend of decreasing with temperature. The trends in the LD parameters follow intuition as well; we expect LD to be more pronounced at lower SOC and lower temperatures, where diffusion processes within the cell are slower. Indeed, the identified $\eta_{th}$ diminishes as we move towards lower SOC and temperatures, expediting the onset of LD. Similarly, both growth parameters. $\theta_{R}$ and $\theta_{\eta}$, are larger in the same region, indicating faster LD resistance growth.

We also evaluate the LUT model performance on the characterization data when fitted with progressively more data. In Fig. \ref{Fig:FitToMPP}(a) we compare the performance of 3 versions of the LUT model; fitted using (1) CCP data only, (2) CCP and FD data, and (3) all characterization data (CCP, FD, and MPP) \footnote{By the time the MPP tests were run, the cells had started to show some resistance growth (see Fig. \ref{Fig:CharacterizationTests}(b)). For this reason, when fitting to MPP, we assumed a uniform 3\% increase in the cell resistance based on the RPT data. This resistance growth is applied to $R_0, R_1, R_2, \theta_{R}$, and $\theta_{\eta}$. This is just an approximation as the aging impact on the resistance is perhaps different at different temperature and SOC combinations. However, as the results in Fig. \ref{Fig:FitToMPP} indicate, this approximation is sufficient to fit the entire characterization data set.}. 
The results indicate that the model trained on CCP data does not generalize very well to deeper discharges seen in the FD data set, nor to the variable current MPP. Similarly, the model does not perform well on the variable current MPP data, when trained on only constant current CCP and FD data sets. This highlights that the CCP data set, commonly used in literature for ECM parameter identification \cite{hentunen2014time, lin2014lumped}, may not be informative enough for high C-rate regimes. Moreover, this observation confirms that the MPP data set indeed contains information that is new for a model built on CCP and FD data sets. Lastly, Fig. \ref{Fig:FitToMPP}(b) shows the model predictions on some MPP examples for the 3 different model versions, confirming the superior performance by the model trained on all characterization data.

\section{Results and Discussion: Model Validation} \label{Sec:Validation}

\subsection{Validation Metric}

Typically, validation of battery models focus on accuracy of predicted voltage. While this can be useful to battery modelers and engineers designing hardware and algorithms, it is not directly relevant to the assessment of flight safety. Joby’s position is that pilots and mission planners will benefit most from the reporting of remaining vehicle capability, best embodied in the form of a “remaining time at condition” statement.  E.g., “the vehicle will have 52 seconds of remaining hover time at the planned destination”. This mirrors the Federal Aviation Administration’s  traditional mandate that every flight plan include a fuel reserve above and beyond what is necessary to reach the intended destination, expressed in terms of time \cite{CFR_part91, CFR_part135}. Therefore, the critical metric for evaluating battery model performance in eVTOL applications is reserve time prediction accuracy. 

Official requirements and associated conditions for eVTOL reserves, as well as indication accuracies, are still under development at the time of this study. Regardless, it is expected that the most challenging reserve prediction task will be predicting at-destination hover reserve during pre-flight planning. This has the longest time horizon and the most challenging, highest power reserve condition. This prediction task shall be used for validation of model performance herein. To this end, we include an extra segment of constant power discharge at the end of each battery power profile to represent the available hover reserve. It is continued until the cell reaches a maximum temperature or minimum voltage limit, and the measured duration is compared to the model predicted duration. The distribution of the differences between these two values will define our model quality.

\subsection{Validation Test Design}

\begin{figure}[t!]
	\begin{center}
		\includegraphics[width=0.9\textwidth]{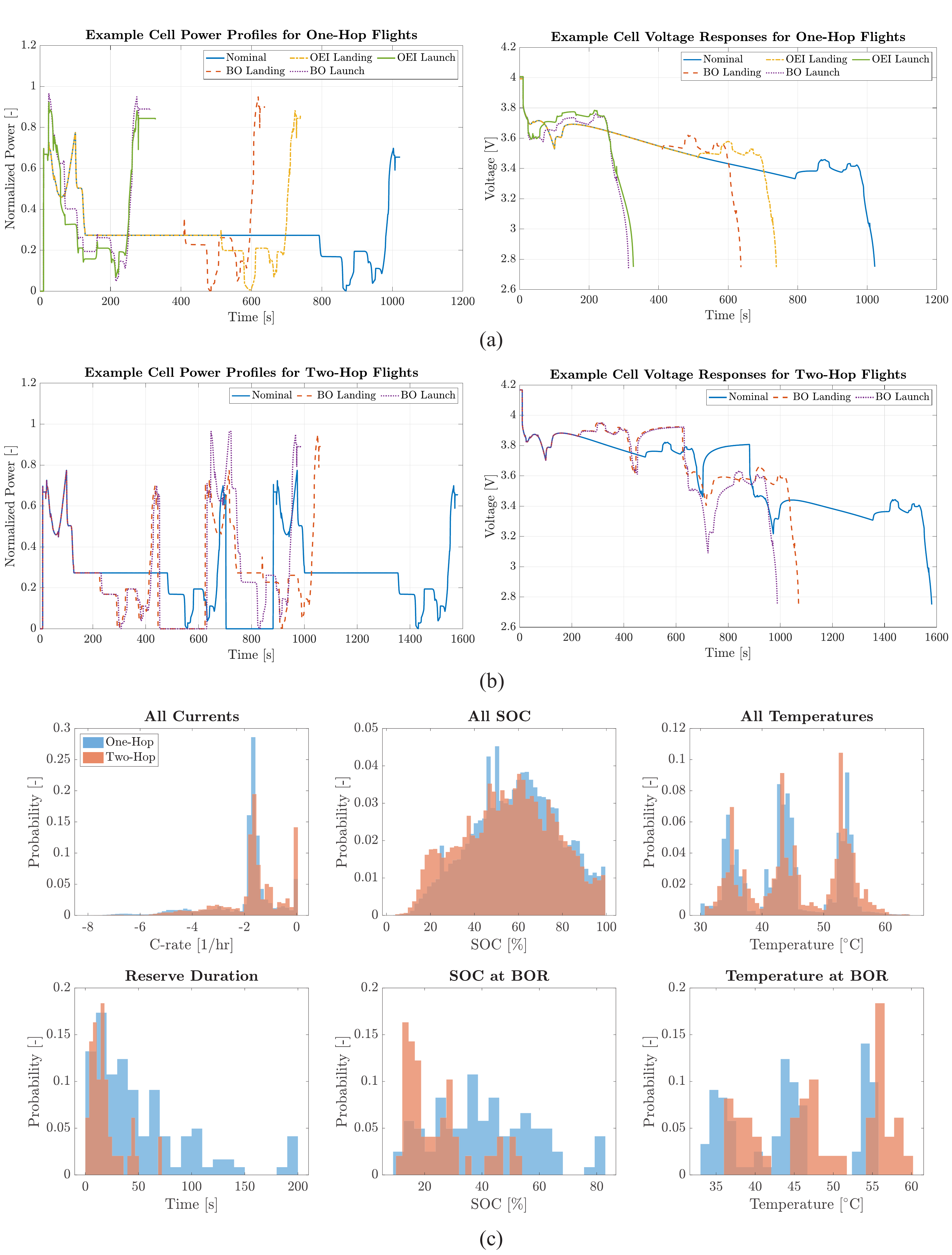}
		\vspace{-0.3cm}
		\caption{Validation flight data set: (a) example power profiles and corresponding cell voltage response for one-hop missions with five different variants, (b) example power profiles and corresponding cell voltage response for two-hop missions with three different variants (other variants are omitted to avoid clutter), and (c) distributions of key metrics through entire flights as well as at the beginning of reserve section for one- and two-hop flights.}
		\vspace{-0.7cm}
		\label{Fig:ValidationData}
	\end{center}
\end{figure}

Our validation test includes a suite of eVTOL flight profiles encompassing a wide array of conditions within the operational envelope. Specifically, we design both one- and two-hop profiles. In a one-hop profile, the vehicle takes off and travels to destination with no stops in between, while a two-hop profile contains two back-to-back flights, with a short rest period of 3 minutes and no charging in between. For each flight, the test set includes a number of possible variants:
\begin{itemize}
	\item Nominal: No faults occur during the flight.
	\item One Engine Inoperable Landing (OEI Landing): A fault case where a single propulsion unit is rendered inoperable at the end of cruise, putting more load on the remaining units, and thereby increasing the power drawn from all battery packs through the descent, transition, and hover landing.
	\item Battery Out Landing (BO Landing): A fault case where one battery pack is rendered inoperable at the end of cruise, increasing the power drawn from the remaining battery packs through the descent, transition, and hover landing.
	\item One Engine Inoperable Launch-Abort (OEI Launch-Abort): A fault case where a propulsion unit is rendered inoperable during takeoff, increasing the power drawn from all battery packs through the rest of the flight including cruise, descent, transition, and hover landing. The cruise altitude and speed are significantly decreased, and the cruise duration is shortened to near zero.
	\item Battery Out Launch-Abort (BO Launch-Abort): A fault case where a single battery pack is rendered inoperable during takeoff, increasing the power drawn from remaining battery packs through the rest of the flight including cruise, descent, transition, and hover landing. The cruise altitude and speed are significantly decreased, and the cruise duration is shortened to near zero.
\end{itemize}
Across flights, we vary vehicle loads such as ambient temperature (which affects propulsion unit power) and vehicle gross takeoff weight. We also examine various battery initial conditions including temperature and SOC. We generate battery power profiles tailored to each combination of flight profiles and test conditions, using a proprietary software package built at Joby. Variations in cruise distance, subject to the constraint of a positive reserve duration, are then included in the validation test set. Overall, this validation set includes a total of 116 one-hop and 49 two-hop flight profiles, each with four replicates. Example power profiles for different variants and the corresponding cell voltage responses are shown in Fig. \ref{Fig:ValidationData}(a-b). Distributions of key metrics on tested power profiles and the battery conditions at beginning-of-reserve (BOR) are given in Fig. \ref{Fig:ValidationData}(c).

\subsection{Model Performance on Validation Data}

\begin{figure}[t!]
	\begin{center}
		\includegraphics[width=0.9\textwidth]{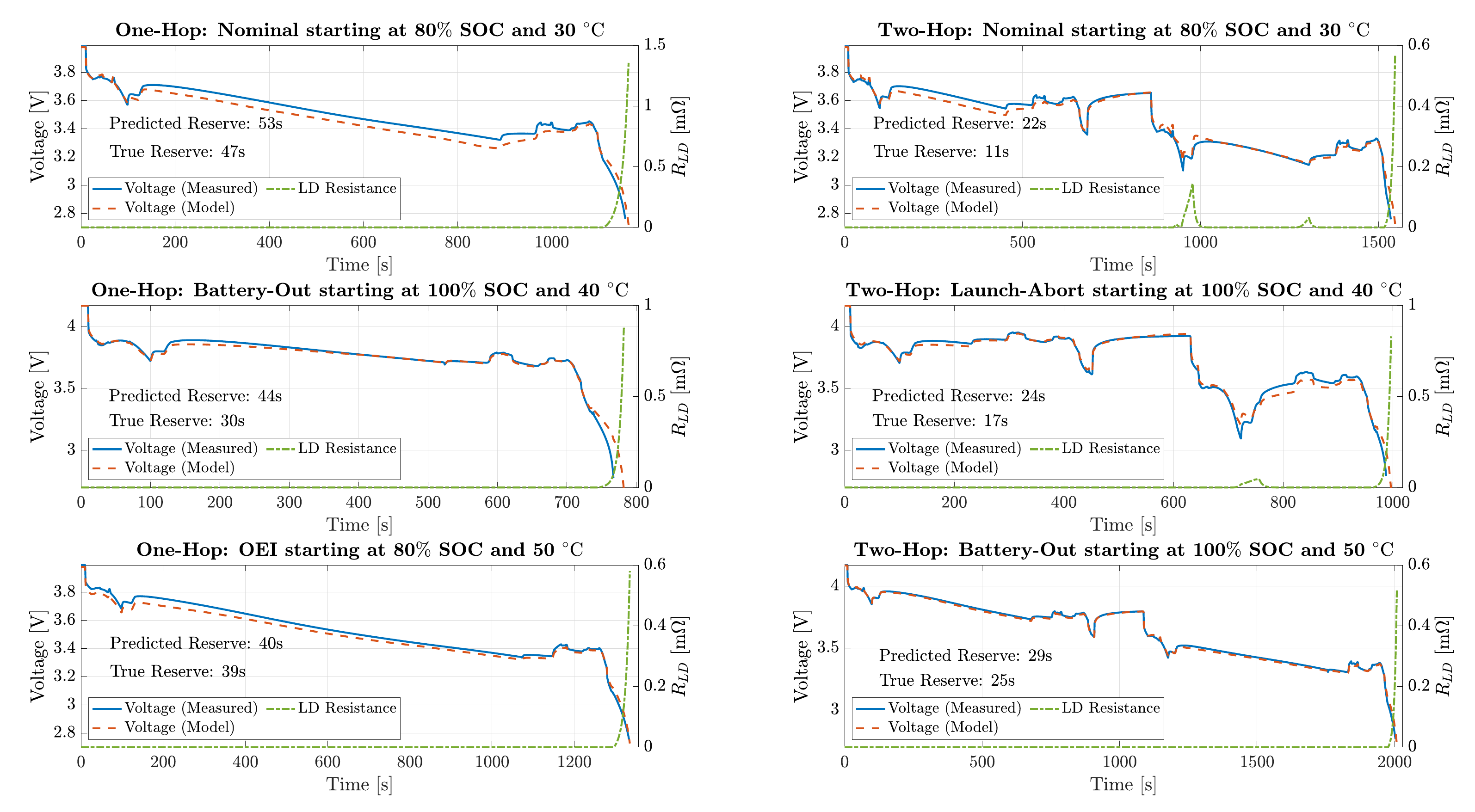}
		\vspace{-0.3cm}
		\caption{Example model performance on several one- and two-hop flights in the validation data set.}
		\vspace{-0.7cm}
		\label{Fig:ValidationResults_Traj}
	\end{center}
\end{figure}

Several examples from one- and two-hop flight data sets at different temperatures are shown in Fig. \ref{Fig:ValidationResults_Traj} along with the corresponding model predictions and the predicted and true reserves for each example. \footnote{Based on the RPT results in Fig. \ref{Fig:CharacterizationTests}(b), the cell resistances were on average $\sim$5\% and $\sim$7\% higher during the one- and two-hop validation flights compared to their respective values during characterization. As such, when evaluating the model on the validation data, we include a 5\% and 7\% uniform resistance growth for one- and two-hop flights, respectively. We also uniformly reduce the threshold overpotential parameter, $\eta_{th}$, by 30 mV to further accommodate the aging impact on the LD resistance growth.}
Overall, the examples show a good match between the model and measurements and the reserve prediction errors are small to moderate in most cases. However, in some cases errors of 100 mV and greater can be observed. This error can be partially attributed to the fact that aging is accounted for by uniformly increasing the resistance parameters across all conditions based on an RPT resistance estimate that is calculated from a 4C pulse at 40 $\rm ^\circ C$. However, the parameters are expected to have changed to a variable degree at different conditions. This in turn highlights the challenge of accounting for aging in LPV ECM formulations. Optimal experimental design methods may be used to address this challenge \cite{rothenberger2015genetic, park2018optimal}, but the application is out of the scope of the present work.

Fig. \ref{Fig:ValidationResults_Traj} also shows the LD resistance values. The LD resistance is only activated under high load scenarios, as intended. This means that in most one-hop flights, the LD resistance only starts to grow near or during the hover reserve at the end of the mission. For some two-hop flights on the other hand, the cell can approach DLC during the takeoff for the second of the two back-to-back flights, where the takeoff SOC is lower due to lack of charging after the first flight.

\begin{figure}[t!]
	\begin{center}
		\includegraphics[width=0.9\textwidth]{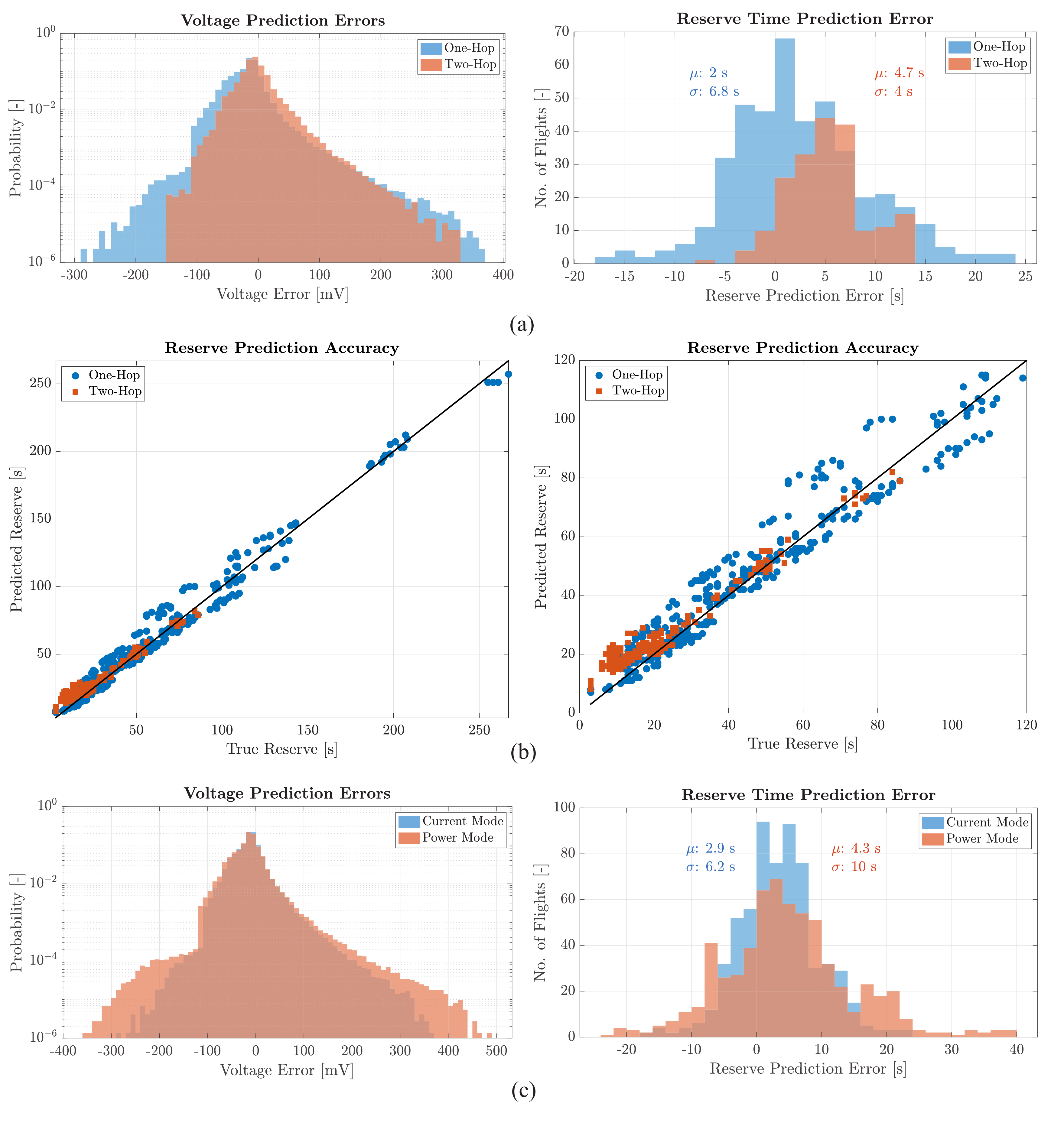}
		\vspace{-0.3cm}
		\caption{Error statistics on all replicates of validation flights: (a) voltage and reserve time prediction error distributions for one- and two-hop flights, (b) predicted vs. true reserve values, where an exploded version is also provided to better highlight the data with true reserve less than 2 minutes, and (c) comparison of voltage and reserve time prediction error distributions for all flights obtained by running the model in current and power modes.}
		\vspace{-0.7cm}
		\label{Fig:ValidationResults_Dist}
	\end{center}
\end{figure}

More comprehensive model prediction error statistics are presented in Fig. \ref{Fig:ValidationResults_Dist}. Specifically, the voltage and reserve prediction error distributions for all replicates of one- and two-hop flights are shown in Fig. \ref{Fig:ValidationResults_Dist}(a), where a positive error indicates model over-prediction. For both sets of flight data, the model voltage error distribution has a slightly longer tail to the right and the reserve prediction errors have a positive mean, indicating a tendency to over-predict the reserve capability. While better accounting for cell aging is expected to improve these statistics, battery algorithms that rely on such a model to estimate the state-of- energy or power (SOE/SOP) can apply a threshold to the prediction to move the distribution to the left and ensure conservative predictions.

A closer look at the reserve prediction errors is provided in Fig. \ref{Fig:ValidationResults_Dist}(b) where a zoomed in version is also shown, highlighting the cases in the data set with less than 2 minutes of true reserve. These cases are more critical, as they are closer to the expected true reserve requirements. The results overall show a good agreement between the model and the measurements. Nonetheless, a slight over-prediction of reserve capability is again observable in this view, especially for the two-hop cases, highlighting the need to better account for aging.

Finally, we note that throughout the model fitting and validation processes so far, we have always used measured current as an input for the model. However, in application, predictions are often made using requested power as input. The model can easily be iterated over to find the current that matches the requested power. However, the process is expected to amplify the model error. To quantify that, we compare the model prediction error distributions for voltage and reserve capability under both current and power mode simulations in Fig. \ref{Fig:ValidationResults_Dist}(c). The figure includes data from all replicates of one- and two-hop flights. As expected, running the model in power mode results in longer tails for the voltage prediction errors. Similarly, the mean of the reserve prediction error distribution is shifted further to the right. 

Overall, these results indicate that accurately capturing a battery cell response in a lithium depleted region can be very challenging. This is due to the fact that the voltage sensitivity to small parameter perturbations grows significantly as the cell gets closer to the DLC. Although the presented enhanced ECM achieves good accuracy on this task, it is not immediately clear if it is sufficient for successful eVTOL operations. Safety margins can be subtracted from the predicted reserve times to accommodate any degree of inaccuracy and achieve a target frequency of conservatism, but larger margins will detract from vehicle utility.


\section{Summary and Conclusions} \label{Sec:Conclusions}

The unique requirements of eVTOL applications lead to challenging conditions for battery packs, including high discharge rates at low SOCs encountered in fault scenarios. With the goal of enabling accurate battery state estimation and performance prediction in said conditions, we provide an enhanced ECM that is derived using data-driven modeling techniques. Specifically, we focus on improving conventional ECM prediction under high C-rate discharge, where diffusion limitations can create a Li$^+$ depleted region in the positive electrode, leading to rapid decrease in cell discharge capacity. 

We show that utilizing a discrepancy modeling framework, a single dynamic resistance (dependent on RC component overpotential) added to a conventional 2RC ECM is sufficient to capture the rapid voltage drop behavior. The dynamics of this model can be represented in a compact state-space form, with only one additional state compared to a 2RC ECM. To harness the full capability of the model architecture, the parameters are allowed to vary with the operating conditions. We provide methods for improved parameter identification using an extensive characterization data set. The identification problem formulation focuses on ensuring smooth parameter tables and avoiding over-fitting with several regularization terms. Results indicate that the constant current pulses commonly used in ECM parameter identification are not sufficient to match cell voltages under eVTOL-like power profiles. Thereby, we created and applied a new eVTOL-focused characterization test method to improve model fits.

The model is validated against a diverse flight data set, measured on large format commercial pouch cells. The flights are designed to cover a wide array of conditions expected for nominal and faulted eVTOL operations. We employ reserve time prediction accuracy as the most relevant metric for assessing model quality, and show that the LD-ECM presented herein can achieve an end-of-mission hover reserve duration prediction error distribution with a standard deviation of 6.2 seconds over all tested validation duty cycles. However, maximum reserve duration prediction errors as large as 20 seconds are observed. This is not unexpected given that the unstable voltage decay behavior is inherently highly sensitive to small parameter perturbations. It is important to highlight that these accuracy metrics may change with time as batteries age depending on how model parameters are varied by the operator through life to account for such effects.

Model accuracy requirements for adequate eVTOL safety have not yet been established, but are the subject of significant ongoing work at Joby. In future publications, we will show how these levels of model accuracy can be managed at the fleet level to yield sufficient fleet safety, along with the value of further improvements to model accuracy.

\newpage
\appendix
\setcounter{table}{0}
\renewcommand{\thetable}{A\arabic{table}}

\section{Functional Representation}\label{Sec:Functional_Forms}

\subsection{Ensuring Non-Negativity}
To satisfy the non-negativity requirement with the functional forms, we use the sum-of-squares (SOS) polynomial representation \cite{packard2010Sos}:
\begin{align}
	\theta_j = \Sigma_{j}(SOC, T, I, \alpha_j), \nonumber
\end{align}
where $\Sigma_{j}$ is the SOS polynomial representing parameter $\theta_{j}$. Constructing an SOS polynomial is straightforward when using quadratic forms; if $z(x)$ is the vector of all monomials of $x$ up to degree $d$, then $\Sigma(x) = z^T(x)LL^Tz(x)$ is an SOS polynomial of degree $2d$, where $L$ is any lower triangular matrix \cite{packard2010Sos}. A similar approach can be used for rational functions, where both the numerator and denominators are chosen to be SOS functions.

\subsection{Parameter Regularization}

Given inherent smoothness in polynomial and rational representations, the regularization cost for functional forms can be simplified:
\begin{align}
	J_{\rm reg}(\boldsymbol{\alpha}) =  w_{\theta_{\eta}}\|\theta_{\eta}\|_1 + w_{\theta_{R}}\|\theta_{R}\|_1 - w_{\eta_{th}}\|\eta_{th}\|_1  - w_{{\tau_2}}\|\tau_{2}\|_1  + w_{reg}\|\boldsymbol{\alpha}\|_2 + w_{constraint}\|C\|_1,  \label{Eq:J_reg_func}
\end{align}
where the first four terms on the right hand side are identical to those in Eq. \ref{Eq:J_reg_lut}, the fifth term is a 2-norm regularization on the coefficient of the function used in fitting, and the last term is the penalty for enforcing soft constraints on parameter upper bounds. Our use of SOS polynomials allows for enforcing the lower bounds by construction (by simply adding the lower bound to the SOS polynomial). This ensures the critical lower bounds that are required for solver stability are never violated\footnote{A negative time constant causes model instability, causing most numerical optimization solvers to crash at that iteration.}. The upper bounds are less critical for numerical solver stability and are implemented as soft constraints through a penalty method \cite{nocedal1999numerical} via the last term in Eq. \ref{Eq:J_reg_func}, leveraging a linear penalty term with a sigmoid activation on parameter values normalized to the unit interval:
\begin{align}
	C_j = \sum_i 0.5\left[1 + \tanh\left(\frac{\theta_{i, normalized}-1}{10^{-3}}\right)\right] \theta_{i, normalized}. \nonumber
\end{align}

\subsection{Lookup Tables vs. Functional Forms: Comparison of Fit Quality on CCP Data}

\begin{figure}[t!]
	\centering
	\begin{subfigure}[t]{0.5\textwidth}
		\centering
		\includegraphics[height=2.2in]{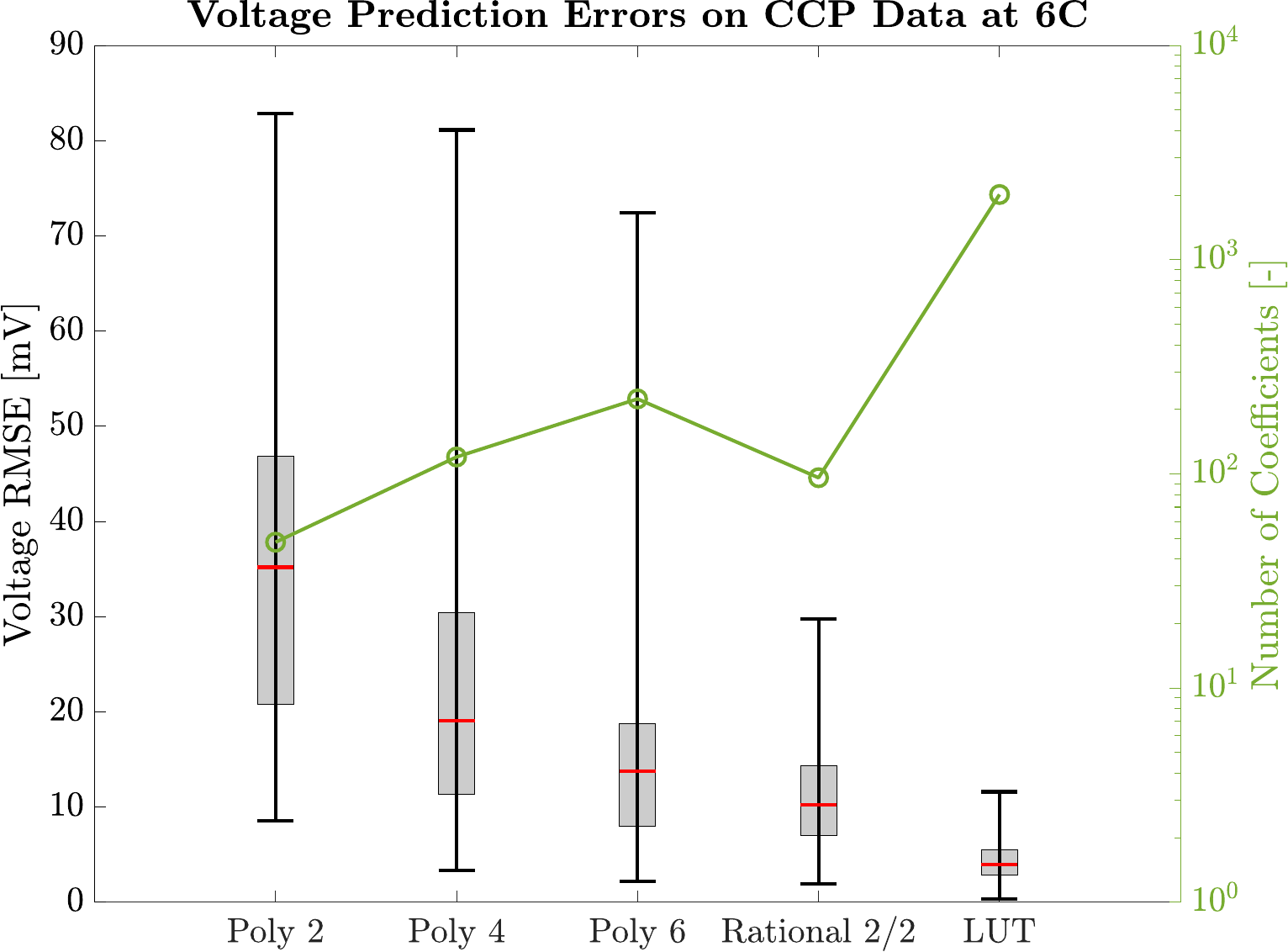}
				\caption{}
	\end{subfigure}%
	~
	\begin{subfigure}[t]{0.5\textwidth}
		\centering
		\includegraphics[height=2.2in]{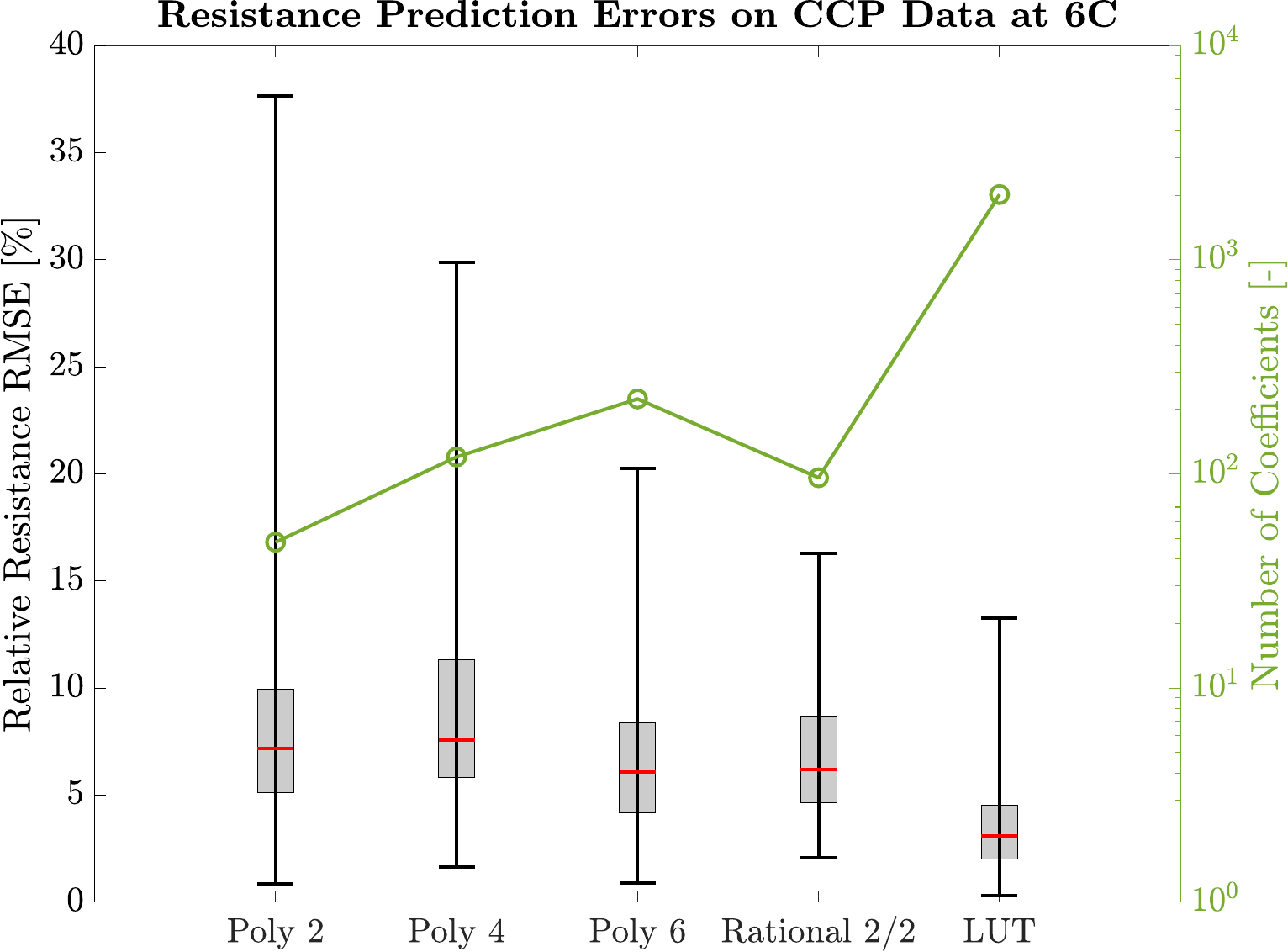}
				\caption{}
	\end{subfigure}
	\caption{Model performance on training data: (a) voltage and (b) ohmic resistance RMSE on 6C CCP data with different parameter representations}
	\label{Fig:LutVsFunctions}
\end{figure}

Fig. \ref{Fig:LutVsFunctions} shows a comparison between the voltage and resistance fit qualities obtained on 6C discharge CCP data with LUT and different functional form representation. The LUT provides the best overall fits, which is not surprising given that the LUT is much more expressive than the functional form and has up to two orders of magnitude more degrees of freedom. Among the functional representations, the average RMSE declines with the polynomial degree, as the representation becomes more expressive. The rational functions clearly outperform the polynomials, as a rational function with quadratic numerator and denominator (12 variables per LD-ECM parameter and 96 total parameters) achieves training errors on par with the LUT. 

The functional forms are often harder to fit; each parameter in the functional form representation can impact all data, whereas the impact of a parameter in the LUT is restricted to a small region around its corresponding grid points. This also has an implication on updating the model with new data. Specifically, when new data are to be considered with the LUT approach, we can only update the LUT entries corresponding to the conditions observed in the new data, whereas with the functional form approach, changing any coefficient in the function would impact the function output everywhere in its domain. 

Fitting a given C-rate from the CCP data usually takes around 5-10 minutes for both the LUT and the low-order polynomials and rational functions. Higher order functions can take significantly longer. While the LUT formulation results in up to two orders of magnitude more decision variables, the resulting Hessian structure is very sparse and the problem can be solved efficiently, whereas the functional form representation results in a dense Hessian. Hence, the solution times for the LUT approach is favorable on bigger data sets or those that contain longer sequences.

For these reasons and its better expressive power, we only used the LUT method to evaluate model performance on validation data. It is conceivable that better motivated functional forms that are potentially constrained by physical laws (e.g., Arrhenius law for variations in the temperature direction), would perform better. But exploration of such functional forms is outside the scope of this work.

\section{Acronyms} \label{Sec:nomenclature}
A complete list of acronyms is provided in Table \ref{Table:Acronyms} below for accessibility.

\begin{center}
	\small
	\begin{longtable}{l l} 
		\caption{List of acronyms.}\\
		\label{Table:Acronyms}
		
		Symbol & Description\\
		\hline
%
%
%

		BMS & Battery management system \\
		BO & Battery out  \\
		BOR & Beginning of reserve \\
		CCP & Constant current pulse \\
		DLC & Diffusion limited C-rate \\
		DOE & Design of experiment \\
		ECM & Equivalent circuit model \\
		eVTOL & Electric vertical takeoff and landing \\
		FD & Full discharge \\
		LD & Lithium depletion \\
		LPV & Linear parameter varying \\
		LUT & Lookup table \\
		MPP & Multi-power pulse \\
		OCV & Open-circuit voltage \\
		OEI & One engine inoperable \\
		RC & Resistor-capacitor \\
		RMSE & Root mean squared error \\
		RPT & Reference performance test \\
		SOC & State of charge \\
		SOE & State of energy \\
		SOH & State of health \\
		SOP & State of power \\
		SOS & Sum of squares \\

	\end{longtable}
\end{center}


%
%
%
%

\bibliographystyle{ieeetr}
\bibliography{ECMref}

\end{document}